\documentclass[letter,10pt]{article} 
\usepackage{color,amssymb,amsthm,amsmath,amsxtra,fullpage,graphicx,hyperref} 

\usepackage{abstract}

\def\bfS{{\textbf{S}}}
\def\mod{\text{ mod }}

\numberwithin{equation}{section}

\begin{document}

\def\bfS{{\textbf{S}}}

\def\todo#1{\textcolor{red}{\textbf{**** TODO -- #1 ****}}}

\renewcommand{\qed}{\nobreak \ifvmode \relax \else
      \ifdim\lastskip<1.5em \hskip-\lastskip
      \hskip1.5em plus0em minus0.5em \fi \nobreak
      \vrule height0.75em width0.5em depth0.25em\fi}

\newtheorem{theorem}{Theorem}[section]
\newtheorem{lemma}[theorem]{Lemma}
\newtheorem{conjecture}[theorem]{Conjecture}
\newtheorem{proposition}[theorem]{Proposition}
\newtheorem{corollary}[theorem]{Corollary}

\theoremstyle{definition}
\newtheorem{example}[theorem]{Example}
\newtheorem{definition}[theorem]{Definition}	
\newtheorem{construction}[theorem]{Construction}

%

\centerline{{\LARGE A Construction for Periodic ZCZ Sequences}}
\medskip
\centerline{\large Samuel T. Blake, Andrew Z. Tirkel}
\centerline{\large\it School of Mathematical Sciences, Monash University, Australia}
\bigskip

\begin{abstract}
	\noindent{\sc Abstract.} We introduce a construction for periodic zero correlation zone (ZCZ) sequences over 
	roots of unity. The sequences 
	share similarities to the perfect periodic sequence constructions of Liu, Frank, and Milewski. The sequences 
	have two non-zero off-peak autocorrelation values which asymptotically approach $\pm 2 \pi$, so the sequences
	are asymptotically perfect.
\end{abstract}

ZCZ sequences see applications in areas including broadband satellite IP networks \cite{Zeng2005}, 
CDMA systems \cite{Suehiro1994}\cite{Fan1999b}, quasi-synchronous code-division multiple-access 
(QS-CDMA) communication systems \cite{DeGaudenzi1992}, and watermarking 
\cite{VanSchyndel2000}\cite{Tirkel2001}. \\

Wolfmann \cite{Wolfmann1992} introduced the idea of ZCZ binary sequences with the so-called 
\textit{almost perfect sequences}, which have one non-zero off-peak 
autocorrelation value. Non-binary ZCZ sequences over 
roots of unity were first constructed by Suehiro \cite{Suehiro1994}. Other types of ZCZ sequences 
have been constructed based on the existence of complementary pairs 
\cite{Suehiro1994}\cite{Fan1999a}\cite{Matsufuji1999}. \\

The periodic cross-correlation of the sequences, $\textbf{a} = \left[a_0,a_1,\cdots, a_{n-1}\right]$ and 
$\textbf{b} = \left[b_0, b_1,\cdots, b_{n-1}\right]$, for shift $\tau$ is given by 
$$\theta_{\textbf{a},\textbf{b}}(\tau) = \sum_{i=0}^{n-1}a_i b_{i+\tau}^*,$$ where $i+\tau$ is computed
modulo $n$. The periodic autocorrelation of a sequence,
\textbf{s} for shift $\tau$ is given by $\theta_{\textbf{s}}(\tau) = \theta_{\textbf{s},\textbf{s}}(\tau)$. For 
$\tau \neq 0 \mod n$, $\theta_{\textbf{s}}(\tau)$ is called an \textit{off-peak} autocorrelation. A sequence 
has good ZCZ autocorrelation if there are a small number of non-zero off-peak autocorrelation values, each of which are 
small in magnitude. Ideally, the non-zero values should be grouped closely together.\\

The periodic autocorrelation of a sequence, $\textbf{s} = [s_0, s_1, \cdots, s_{ld^2-1}]$, can be expressed in terms of the autocorrelation and 
cross-correlation of an array {\it associated} with \textbf{s} \cite{Heimiller1961}\cite{Frank1962}\cite{Mow1993}.  The sequence 
\textbf{s} has the 
{\it array orthogonality property} (AOP) for the {\it divisor} $d$ if the array 
\textbf{S} associated with \textbf{s} has the following two properties:
\begin{enumerate}
\item For all $\tau$, the periodic cross-correlation of any two distinct columns of \textbf{S} is zero.
\item For $\tau\neq0$, the sum of the periodic autocorrelation of all columns of \textbf{S} is zero. 
\end{enumerate}

Any sequence with the AOP is perfect \cite{Mow1993}. It is possible to relate the non-zero autocorrelation values of a sequence
to the non-zero autocorrelation or cross-correlation values in the two conditions of the AOP. A non-zero in 
the autocorrelation for shift $\tau = q'd+r'$, ($r'<d$), corresponds to a non-zero in the 
cross correlation for shift $\kappa = q' + \left\lfloor\frac{r+r'}{d} \right\rfloor$, ($0\leq r < d$) for the first condition 
of the AOP($r'\neq0$) or a non-zero in the sum of the autocorrelations for shift 
$\kappa = q' + \left\lfloor\frac{r}{d} \right\rfloor$, ($0\leq r < d$) for the second condition of the AOP ($r'=0$). \\

In most perfect sequence constructions, one proves the sequence has perfect autocorrelation by reducing the 
autocorrelation to a Gaussian summation. A Gaussian summation is given by 
$\sum_{k=0}^{n-1}\omega^{q k}$, where $\omega = e^{2 \pi \sqrt{-1}/n}$ and $q \in \mathbb{Z}$. 
If $q \neq 0 \mod n$, then the sum is zero. \\

We now introduce a construction for ZCZ sequences over roots of unity. This construction has similarities to 
the perfect periodic sequence construction of Liu \cite{Liu2004} in that it uses the {\it floor} function within its index 
function.\footnote{Note that our construction, like the construction of Liu, also holds for the {\it ceiling} function.} Furthermore, 
it is similar to the perfect periodic sequence 
constructions of Frank and Milewski \cite{Frank1962}\cite{Milewski1983} in that it is formed by concatenating, 
row-by-row, a perfect array. Furthermore, this construction is very similar to the perfect sequence construction by the 
authors \cite{Blake2012}. \\

Let \textbf{s} be a sequence of length $24(2n+1)$ over $6(2n+1)$ roots of unity, where $n \in \mathbb{N}$. Construct a 
$12(2n+1)\times2$ array, \textbf{S}, over $6(2n+1)$ roots of 
unity, where $\textbf{S} = [S_{i,j}] = \omega^{\left\lfloor i (i + j)/2\right\rfloor}$ and $\omega = e^{2 \pi \sqrt{-1}/(6(2n+1))}$. 
The sequence, \textbf{s} is constructed by enumerating, row-by-row, the array \textbf{S}.\\

We show \textbf{s} has periodic ZCZ autocorrelation. We begin by showing \textbf{s} 
satisfies the first condition of the AOP. That is, we show that the cross-correlation of the two columns of \textbf{S} is 
zero for every shift, $\kappa$. The cross-correlation of the two columns of \textbf{S} is given by

\begin{align*}
\theta_{S_{i,0}, S_{i,1}}(\kappa) &= \sum_{i=0}^{12(2n+1)-1} S_{i,0} S_{i+\kappa,1}^*\\
&= \sum_{i=0}^{12(2n+1)-1} \omega^{\left\lfloor \frac{i^2}{2} \right\rfloor} 
		\omega^{-\left\lfloor \frac{(i+\kappa)^2 + i + \kappa}{2} \right\rfloor}\\
&=  \sum_{i=0,2,4,\cdots}^{12(2n+1)-1} \omega^{\left\lfloor \frac{i^2}{2} \right\rfloor} 
		\omega^{-\left\lfloor \frac{(i+\kappa)^2 + i + \kappa}{2} \right\rfloor}
	+ \sum_{i=1,3,5,\cdots}^{12(2n+1)-1} \omega^{\left\lfloor \frac{i^2}{2} \right\rfloor} 
		\omega^{-\left\lfloor \frac{(i+\kappa)^2 + i + \kappa}{2} \right\rfloor}\\
&=  \sum_{k=0}^{6(2n+1)-1} \omega^{\left\lfloor \frac{(2k)^2}{2} \right\rfloor} 
		\omega^{-\left\lfloor \frac{(2k+\kappa)^2 + 2k + \kappa}{2} \right\rfloor}
	+ \sum_{k=0}^{6(2n+1)-1} \omega^{\left\lfloor \frac{(2k+1)^2}{2} \right\rfloor} 
		\omega^{-\left\lfloor \frac{(2k+1+\kappa)^2 + 2k+1 + \kappa}{2} \right\rfloor}\\
&= \omega^{\left\lfloor\frac{\kappa^2+\kappa}{2}\right\rfloor} \left(1+\omega^{-\kappa-1}\right) 
		\sum_{k=0}^{6(2n+1)-1} \omega^{-(2\kappa+1)k}.
\end{align*}
The (Gaussian) summation above is zero as $-2\kappa-1 \neq 0 \mod 6 (2n+1)$. Thus, \textbf{s} satisfies the first condition of the AOP, which 
implies the autocorrelation of the sequence, \textbf{s} is zero for $\tau \neq 0 \mod 2$. \\

We now show \textbf{s} satisfies the second condition of the AOP with the exception of two off-peak shifts. That 
is, we show that the sum of the autocorrelation of the two columns of \textbf{S} is zero for all but two shifts. 

\begin{equation}
\theta_{S_{i,0}}(\kappa) + \theta_{S_{i,1}}(\kappa) =  \sum_{i=0}^{12(2n+1)-1} S_{i,0} S_{i+\kappa,0}^* + 
	 \sum_{i=0}^{12(2n+1)-1} S_{i,1} S_{i+\kappa,1}^* \tag*{(1) + (2)}
\end{equation}
From which (1) becomes  
\begin{align*}
&\sum_{i=0}^{12(2n+1)-1} \omega^{\left\lfloor\frac{i^2}{2}\right\rfloor} 
	\omega^{-\left\lfloor\frac{(i+\kappa)^2}{2}\right\rfloor}\\
&=  \sum_{i=0,2,4,\cdots}^{12(2n+1)-1} \omega^{\left\lfloor\frac{i^2}{2}\right\rfloor} 
	\omega^{-\left\lfloor\frac{(i+\kappa)^2}{2}\right\rfloor} + 
	\sum_{i=1,3,5,\cdots}^{12(2n+1)-1} \omega^{\left\lfloor\frac{i^2}{2}\right\rfloor} 
	\omega^{-\left\lfloor\frac{(i+\kappa)^2}{2}\right\rfloor} \\
&= \sum_{k=0}^{6(2n+1)-1} \omega^{\left\lfloor\frac{(2k)^2}{2}\right\rfloor} 
	\omega^{-\left\lfloor\frac{(2k+\kappa)^2}{2}\right\rfloor} + 
	\sum_{k=0}^{6(2n+1)-1} \omega^{\left\lfloor\frac{(2k+1)^2}{2}\right\rfloor} 
	\omega^{-\left\lfloor\frac{(2k+1+\kappa)^2}{2}\right\rfloor}\\
&= \left(\omega^{-\left\lfloor\frac{\kappa^2}{2}\right\rfloor} + 
			\omega^{-\left\lfloor\frac{(1+\kappa)^2}{2}\right\rfloor}\right)
		\sum_{k=0}^{6(2n+1)-1} \omega^{-2\kappa k}.
\end{align*}
Similarly, (2) becomes 
\begin{align*}
&\sum_{i=0}^{12(2n+1)-1} \omega^{\left\lfloor\frac{i^2+i}{2}\right\rfloor} 
	\omega^{-\left\lfloor\frac{(i+\kappa)^2 + i + \kappa}{2}\right\rfloor}\\
&=\sum_{i=0,2,4,\cdots}^{12(2n+1)-1} \omega^{\left\lfloor\frac{i^2+i}{2}\right\rfloor} 
	\omega^{-\left\lfloor\frac{(i+\kappa)^2 + i + \kappa}{2}\right\rfloor} + 
\sum_{i=1,3,5,\cdots}^{12(2n+1)-1} \omega^{\left\lfloor\frac{i^2+i}{2}\right\rfloor} 
	\omega^{-\left\lfloor\frac{(i+\kappa)^2 + i + \kappa}{2}\right\rfloor}\\
&=\sum_{k=0}^{6(2n+1)-1} \omega^{\left\lfloor\frac{(2k)^2+2k}{2}\right\rfloor} 
	\omega^{-\left\lfloor\frac{(2k+\kappa)^2 + 2k + \kappa}{2}\right\rfloor} + 
\sum_{k=0}^{6(2n+1)-1} \omega^{\left\lfloor\frac{(2k+1)^2+2k+1}{2}\right\rfloor} 
	\omega^{-\left\lfloor\frac{(2k+1+\kappa)^2 + 2k+1 + \kappa}{2}\right\rfloor}\\
&= \left(\omega^{-\left\lfloor\frac{\kappa^2+\kappa}{2}\right\rfloor} + 
	 \omega^{-\left\lfloor\frac{\kappa^2+3\kappa}{2}\right\rfloor}\right) \sum_{k=0}^{6(2n+1)-1} \omega^{-2\kappa k}.
\end{align*}

Thus, 

\begin{align*}
\theta_{S_{i,0}}(\kappa) + \theta_{S_{i,1}}(\kappa) &= (1) + (2)\\
&= \left(\omega^{-\left\lfloor\frac{\kappa^2}{2}\right\rfloor} + 
			\omega^{-\left\lfloor\frac{(1+\kappa)^2}{2}\right\rfloor} + 
			\omega^{-\left\lfloor\frac{\kappa^2+\kappa}{2}\right\rfloor} + 
	 		\omega^{-\left\lfloor\frac{\kappa^2+3\kappa}{2}\right\rfloor} \right)
	 	\sum_{k=0}^{6(2n+1)-1} \omega^{-2\kappa k}.
\end{align*}

This (Gaussian) sum is non-zero when $\kappa = 3(2n+1)$, or $\kappa = 6(2n+1)$, or $\kappa = 9(2n+1)$. When $\kappa = 6(2n+1)$, 
$\omega^{-\left\lfloor\frac{\kappa^2}{2}\right\rfloor} + 
			\omega^{-\left\lfloor\frac{(1+\kappa)^2}{2}\right\rfloor} + 
			\omega^{-\left\lfloor\frac{\kappa^2+\kappa}{2}\right\rfloor} + 
	 		\omega^{-\left\lfloor\frac{\kappa^2+3\kappa}{2}\right\rfloor} = 0$. So \textbf{s} satisfies the second condition of the AOP
at all non-zero shifts except $\kappa = 3(2n+1)$ and $\kappa = 9(2n+1)$.\\  

Thus, the autocorrelation of \textbf{s} has two non-zero values at shifts $\tau = 6(2n+1)$ and $\tau = 18(2n+1)$. The 
two non-zero autocorrelation values are equal, and are given by 
$$(-1)^{n+1} 12 (2n+1)\sin\left(\frac{\pi}{6 (2n+1)}\right),$$
which asymptotically approaches $2\pi$ for odd $n$ and $-2\pi$ for even $n$. As a result, for arbitrarily large lengths, 
the ratio of the non-zero off-peak autocorrelation values and the length approaches zero. So, in a sense, \textbf{s} is an
asymptotically perfect sequence. 

\bibliographystyle{abbrv}

\end{document}